\newcommand{\dr}{\textup{d}}

\newcommand{\edot}{\dot{E}}
\newcommand{\edotsev}{\dot{E}_{\Delta}}

\newcommand{\henon}{H{\'e}non}

\newcommand{\rh}{r_{\rm h}}

\newcommand{\mdot}{\dot{M}}

\newcommand{\mmean}{\langle{m}\rangle}

\newcommand{\phil}{\phi_\Delta}
\newcommand{\phim}{\langle\phi\rangle}

\newcommand{\tms}{\tau_{\rm ms}}
\newcommand{\tdf}{\tau_{\rm df}}
\newcommand{\trh}{\tau_{\rm rh}}

\newcommand{\vsql}{v^2_\Delta}

\newcommand{\vsqm}{\langle v^2 \rangle}

\def\apj{ApJ}%          % Astrophysical Journal
\def\apjl{ApJ}%          % Astrophysical Journal, Letters
\def\apjs{ApJS}%          % Astrophysical Journal, Supplement
\def\aap{A\&A}%          % Astronomy and Astrophysics
\def\mnras{MNRAS}%          % Monthly Notices of the RAS
\def\na{New A}%          % New Astronomy
\documentclass[11pt,twoside]{article}
\usepackage{asp2010}

\resetcounters

\begin{document}

\title{Mass loss of stars in star clusters: an energy source for dynamical evolution}
\author{Mark Gieles}
\affil{Institute of Astronomy,  University of Cambridge, Madingley Road, Cambridge CB3 0HA}

\begin{abstract}
Dense star clusters expand until their sizes are limited by the tidal
field of their host galaxy. During this expansion phase the member
stars evolve and lose mass. We show that for clusters with short
initial relaxation time scales ($\lesssim$100 Myr) the dynamical
expansion is largely powered by mass loss from stars in the core, but
happens on a relaxation time scale. That is, the energy release
following stellar mass loss is in balance with the amount of energy
that is transported outward by two-body relaxation.
\end{abstract}

%%%%%%%%%%%%%%%%%%%%%%%%%%%%%%%%%%%%%%%%%%%%%%%%%%%%%%%%%%%%%%%
\section{Introduction}
Like a star, a stellar cluster produces heat by contracting.  The
positive energy that is released by the more tightly bound core is
diffused outward by two-body relaxation \citep{1980MNRAS.191..483L} on
a relaxation time scale
\citep[equation~\ref{eq:trh},][]{1971ApJ...164..399S}.  Just as for a
star, the contraction phase in a cluster is only a small fraction of
the total lifetime\footnote{This is true for clusters that are
  initially much denser than the density within the tidal radius.}.
\citet{1975IAUS...69..133H} had the insight that the evolution of
clusters after core contraction can be understood by assuming that an
energy source provides the right amount of energy in a self-regulating
fashion to sustain the two-body relaxation process.  A breakthrough in
the understanding of cluster evolution, comparable to Eddington's work
on stellar structure.

In single-mass clusters the energy is supplied by binary stars
\citep{1961AnAp...24..369H} that form in the densest phase of core
collapse \citep{2012NewA...17..272T}. They provide energy by
interacting with other stars, thereby increasing the total (negative)
energy of the cluster \citep{1975MNRAS.173..729H}.  In more realistic
clusters mass loss of member stars also increases the total (negative)
energy. In this contribution we show that the corresponding energy
production rate can come into balance with the flow of energy due to
two-body relaxation.
  
  %%%%%%%%%%%%%%%%%%%%%%%%%%%%%%%%%%%%%%%%%%%%
\section{Time scales}
\label{sec:analytic}
When considering mass loss of stars in a collisional system there are
a few time scales that need to be considered: the timescale for
massive stars to segregate to the cluster core by dynamical friction,
and the lifetime of stars, or main sequence time, $\tms$. The former
depends on the mass of the star, $m$, the mean stellar mass, $\mmean$
and the half-mass relaxation time scale of the cluster, $\trh$, as
$\tdf\simeq(m/\mmean)^{-1}\trh$ \citep{1969ApJ...158L.139S}. We use
the conventional expression for $\trh$ \citep{1971ApJ...164..399S}
\begin{equation}
\trh=0.138\frac{N^{1/2}\rh^{3/2}}{\sqrt{G\mmean}\ln\Lambda}.
 \label{eq:trh}
\end{equation}
Here $N$ is the number of stars, $\rh$ is the half-mass radius, $G$ is
the gravitational constant and $\ln\Lambda$ is the Coulomb
logarithm. The argument of the Coulomb logarithm is
$\Lambda\simeq0.11N$ for single-mass systems
\citep{1994MNRAS.268..257G} and $\Lambda\simeq0.02N$ for clusters with
a wide mass spectrum \citep{1996MNRAS.279.1037G}.  The mass-loss rate
is predominantly set by $\tms$ and this timescale depends on the
properties of the star, mainly $m$, and not on the dynamical state of
the cluster.  The two timescales $\tdf$ and $\tms$ become comparable
for the most massive main sequence star after some period of
time. This is because $\tms\propto m^{-\lambda}$, with
$2.5\lesssim\lambda\lesssim3$, and $\tdf\propto m^{-1}$. A star for
which $\tdf(m)=\tms(m)$ reaches the core within its lifetime and will
lose its mass there.  From that moment onwards mass loss will proceed
on average from the central parts of the cluster.  The amount of
energy that is released depends on the mass and specific energy of the
star at the moment it loses mass. In \S~\ref{sec:demand} we quantify
how much energy a cluster needs to evolve and in \S~\ref{sec:supply}
we show how much energy can be generated by mass loss.

%____________________________________________________________________
\section{Energy demand}
\label{sec:demand}
Let us first consider how much energy is needed per unit of time.
Just as for a star, the flow of energy, or luminosity $\edot$, is
determined by the properties of the system as a whole. It equals a
fraction $\zeta$ of the instantaneous total energy $E$ per unit of
$\trh$ \citep{1961AnAp...24..369H,2003gmbp.book.....H,
  2011MNRAS.413.2509G}
\begin{equation}
\frac{\edot}{E}=-\frac{\zeta }{\trh}.
\label{eq:edot}
\end{equation}
The total energy for a cluster in virial equilibrium may be written as
\begin{equation}
E = -\kappa\frac{GM^2}{\rh},
\label{eq:e}
\end{equation}
where $\kappa\simeq0.20-0.25$, depending on the density profile and
$M=\mmean N$ is the total mass. The evolution of the main cluster
parameters ($N$ and $\rh$) can be understood once we know how $\trh$
evolves.  For this it is insightful to consider the idealised models
of \henon.  In the absence of a tidal field the energy flow leads to a
self-similar expansion at a roughly constant mass and $\trh$ grows
linearly in time such that $\rh \propto t^{2/3}$
\citep{1965AnAp...28...62H}.  In a steady tidal field the radius is
limited by the tidal radius and $\trh$ decreases linearly in time
\citep[][]{1961AnAp...24..369H}.  In both cases $\zeta\simeq0.1$ (see
the derivation from \henon 's work in \citealt{2011MNRAS.413.2509G}
and see \citealt{alexander2012} for measurements of $\zeta$ from
numerical simulations) and $\zeta$ can be more than an order of
magnitude higher for clusters with a wide mass spectrum
\citep{2010MNRAS.408L..16G}.

We consider clusters that are not strongly confined by a tidal
field. This is justified by the observation that about half of the
(old) Galactic globular clusters are still expanding towards their
tidal boundary \citep{2010MNRAS.401.1832B, 2011MNRAS.413.2509G}. We
then find from equation~(\ref{eq:trh}) and (\ref{eq:edot}) that
$\trh=(3/2)\zeta t$ and

\begin{equation}
\frac{\edot}{E} = -\frac{2/3}{t}.
\label{eq:edott}
\end{equation}
This means that the (absolute value of the) cluster energy decreases
by roughly $80\%$ each age dex during the expansion.  It is in this
phase that stellar mass loss can provide the energy. Although a
cluster only loses about half of its initial mass due to stellar mass
loss over a Hubble time, this modest mass-loss rate in fact generates
energy at a rate comparable to what is required
(equation~\ref{eq:edott}), as we will show in \S~\ref{sec:supply}.

%____________________________________________________________________
\section{Energy supplied by mass loss}
\label{sec:supply}
As we aim to show in this section, the rate of energy increase as a
result of mass loss of stars, $\edotsev$, leads to a power-law decline
of $-E$ (equation~\ref{eq:edott}). This is primarily because the
evolution of the total mass as a result of stellar evolution from a
stellar system with a Salpeter like stellar initial mass function,
declines roughly as a power of time $M(t)\propto t^{-\nu}$, with
$\nu\simeq0.07$ \citep[this corresponds to a loss of about 15\% of the
  mass every age dex, e.g.][]{2003MNRAS.344.1000B}. The instantaneous
mass-loss rate is thus
 \begin{equation}
 \frac{\mdot}{M}=-\frac{\nu	}{t}.
\label{eq:mdot}
 \end{equation}

To relate this to energy we need to know from where and how fast mass
is lost. We assume that the mass-loss time scale is much longer than
the local crossing time, such that the cluster responds adiabatically
and retains virial equilibrium. \citet{1980ApJ...235..986H} discussed
the adiabatic response of a gravitational system if mass is lost
homologous with the density profile of the cluster, that is, without a
preferred location. The cluster radius then increases as $M^{-1}$ and
the change in energy relates to the change in mass as
(equation~\ref{eq:e})

\begin{equation}
\frac{\dr E}{E} = 3\frac{\dr M}{M}.
\label{eq:dehom}
\end{equation} 

When $\tdf<<\tms$ mass loss happens predominantly from the core of the
cluster, because the most massive stars, i.e. the ones contributing
most to the total stellar mass loss, are centrally concentrated
because of dynamical friction\footnote{This simple view can get
  slightly more complicated if the turn-off mass is significantly
  lower than the non-evolving remnants. If many remnants are retained
  they will eventually dominate the dynamics in the inner regions of
  cluster.}.  Mass loss affects both the potential energy $U=2E$ and
the kinetic energy $T=(1/2) M\vsqm$, where $\vsqm$ is the mean-square
velocity of the system. This is because the mass that is removed
carries away a bit of potential and kinetic energy and because of the
subsequent redistribution of potential and kinetic energy of the
remaining stars.  We can express the change in the potential energy in
terms of the specific potential of the mass that is lost, $\phil$,
relative to the average specific potential of all the stars,
$\phim=2U/M$, as $\dr U/U =2(\phil/\phim)\dr M/M$.  For the kinetic
term we need to compare the squared velocity of the mass that is lost,
$\vsql$, to the mean square velocity of the cluster: $\dr T/T =
(\vsql/\vsqm) \dr M/M$. Using our assumption that the remaining stars
quickly restore virial equilibrium ($E=U/2=-T$) we can relate the
change in the total energy to the change in mass

\begin{eqnarray}
\frac{\dr E}{E}&= &2\frac{\dr U}{U} - \frac{\dr T}{T},\\
            &=& x\frac{\dr M}{M}, \hspace{1.5cm}\mbox{with}\hspace{0.1cm}x\equiv\ 4\frac{\phil}{\phim}-\frac{\vsql}{\vsqm}.
\label{eq:exm}
\end{eqnarray}
For homologous mass loss $\langle\phil\rangle=\phim$ and
$\langle\vsql\rangle=\vsqm$ and therefore $x=3$. This is Hills' result
(equation~\ref{eq:dehom}).  The general result for the rate of energy
`supply' is

\begin{equation}
\frac{\edotsev}{E}  = -\frac{x\nu}{t}.
\label{eq:xnu}
\end{equation}

From a comparison to the energy `demand' (equation~\ref{eq:edott}) we
see that {\it a balance between energy supply and demand is reached if
  $x\nu\simeq2/3$.}  Because $\nu\simeq0.07$ the condition for energy
generation by stellar mass loss to be in balance by two-body
relaxation becomes $x\simeq 10$. To answer the question whether this
value is realistic, we need to consider a cluster potential that
described a cluster in the balanced evolution phase. One example is
the \citet{1983MNRAS.202..995J} model, which has the required $r^{-2}$
density cusp.  From the potential of this model we find
$\phi(r)/\phim=\ln(1+\rh/r)$. If we ignore the small contribution of
the kinetic term to $x$ in equation (\ref{eq:exm}) we find that
$x\simeq 10$ if mass loss takes place at $r\simeq0.1\rh$. This is
probably realistic in the case of a mass segregated system.  This
shows that the product $x\nu$ can be of sufficient magnitude (about
$2/3$, equation~\ref{eq:edott}) to drive the relaxation process. In a
forthcoming paper (Gieles, Heggie \& Church) we will illustrate this
result with numerical simulations and show that stellar mass loss is
indeed a viable energy source in driving the dynamical expansion of
star clusters.

\acknowledgements MG thanks Douglas Heggie for many interesting
discussions on this topic and the staff of the Sterrekundig Instituut
Utrecht for their inspiration and support during the 10 years he spent
there as an undergraduate and PhD student.

\end{document}